%% file: luca.tex
\documentstyle[12pt,agums]{article}
\newcommand{\gra}[1]{\mbox{\boldmath $ #1 $}}

\lefthead{SORRISO-VALVO ET AL.}
\righthead{PHOTOSPHERIC MAGNETIC PRELUDE TO FLARES}
\input{epsf}

\cpright{AGU}{1998}

\ccc{0148-0227/98/98JA-00000\$09.00}

\authoraddr{Valentina~I.~Abramenko,           
Crimean Astrophysical Observatory, 98409 Nauchny, Crimea, Ukraine.
(e-mail: avi@crao.crimea.ua)}   

\authoraddr{Vincenzo Carbone, Luca Sorriso-Valvo and Pierluigi Veltri,
Dipartimento di Fisica and Istituto Nazionale di Fisica per la 
Materia, Ponte P. Bucci, Cubo 31C, Universit\'a della Calabria, I-87036 
Rende (CS), Italy.
(e-mail: carbone@fis.unical.it; sorriso@fis.unical.it; veltri@fis.unical.it)}

\authoraddr{Alain~Noullez and Helene~Politano,
Laboratoire Cassini, Observatoire de la C\^ote d'Azur, 
B.~P.~4229, F-06304 Nice Cedex 04, France.
(e-mail: anz@obs-nice.fr; politano@obs-nice,fr)}

\authoraddr{Annick~Pouquet, ASP/NCAR, P.O. Box 3000 Boulder, 
CO 80307-3000. (e-mail: pouquet@ucar.edu)}

\authoraddr{V.~Yurchyshyn, 
Big Bear Solar Observatory, 40386 North Shore Lane, 
Big Bear City, CA 92314-9672, U.S.A. 
(e-mail: vayur@bbso.njit.edu )}   

\input{update.tex}

\begin{document}

\title{Topological changes of the photospheric magnetic field inside active 
regions: a prelude to flares}

\author{L.~Sorriso-Valvo\altaffilmark{1,2}, V.~Carbone\altaffilmark{1}, 
V.~Abramenko\altaffilmark{3}, V.~Yurchyshyn\altaffilmark{3,4}, 
A.~Noullez\altaffilmark{2}, H.~Politano\altaffilmark{2}, 
A.~Pouquet\altaffilmark{5}, P.~Veltri\altaffilmark{1}} 

\altaffiltext{1}{Dipartimento di Fisica, Universit\'a della Calabria, and 
Istituto Nazionale di Fisica della Materia, sezione di Cosenza, Italy.}
\altaffiltext{2}{Observatoire de la C\^ote d'Azur, Nice, France}
\altaffiltext{3}{Crimean Astrophysical Observatory, Nauchny, Crimea, Ukraine}
\altaffiltext{4}{Big Bear Solar Observatory, Big Bear City, CA}
\altaffiltext{5}{ASP/NCAR, Boulder, CO}

\newpage
\begin{abstract}
The observations of magnetic field variations as a signature of flaring 
activity is one of the main goal in solar physics. Some efforts in the past 
give apparently no unambiguous observations of changes. We observed that the 
scaling laws of the current helicity inside a given flaring active region 
change clearly and abruptly in correspondence with the eruption 
of big flares at the top of that active region. Comparison with numerical 
simulations of MHD equations, indicates that the change of scaling behavior 
in the current helicity, seems to be associated to a topological reorganization
of the footpoint of the magnetic field loop, namely to dissipation of small 
scales structures in turbulence. It is evident that the possibility of 
forecasting in real time high energy flares, even if partially, has a wide 
practical interest to prevent the effects of big flares on Earth and its 
environment.   
\end{abstract}

\newpage
\begin{article}

Solar flares are sudden, transient energy release above active regions of the 
Sun (Priest,~1982). As a consequence of random motion of the footpoints of 
the magnetic field in the photospheric convection (Parker,~1988), flares 
represent the dissipation at the many tangential discontinuities arising 
spontaneously in the magnetic fields of active regions. The magnetic energy is 
released in various form as thermal, soft and hard X-ray, accelerated 
particles etc. The observations of magnetic field variations, as a signature 
of flares above active regions, has been one of the main goals in solar 
physics, and some attempts for this has been made in the past 
({\it e.g.} Hagyard~{\em et~al.},~1999 and references therein). 
All efforts give apparently no unambiguous 
observations of changes. This is due to the fact that often investigations 
try to look for changes of the vector magnetic field as a whole. Recently, 
unambiguous observations of changing have been reported by
Yurchyshyn~{\em et~al.},~2000. The authors observed some typical changes of 
the scaling behavior of the current helicity calculated inside an active 
region of the photosphere, connected to the eruption of big flares above 
that active region. In the present paper we conjecture that the changes in 
the scaling behavior of the observed quantity is related to the occurrence 
of changes in the topology of the magnetic field at the footpoint of the loop. 
 
The occurrence of scaling of signed measures calculated from scalar fields
$f(\gra x)$ which oscillate in sign can be studied through the  following
steps. First of all, introduce the signed measure
   \begin{equation} 
   \mu_i(r) = \int_{Q_i(r)} f(\gra x) d\gra x
   \label{mu} 
   \end{equation} 
through a coarse-graining of non overlapping boxes $Q_i(r)$   of size $r$, 
covering the whole field  defined on a region of size $L$. 
It has been observed (Ott~{\em et~al.},~1992) that, for fields presenting 
self-similarity, this quantity displays well defined scaling laws. That is, 
in a range of scales~$r$, the partition function $\chi(r)$, defined as
   \begin{equation} 
   \chi(r) = \sum_{Q_i(r)} |\mu_i(r)| \sim r^{-\kappa}
   \label{chi} 
   \end{equation} 
where the sum is extended over all boxes occurring at a given scale $r$,
follows a power-law behavior
   \begin{equation} 
   \chi(r) \sim r^{-\kappa}\, .
   \label{chilaw} 
   \end{equation} 
The  scaling exponent $\kappa$ has been called
cancellation exponent (Ott {\em et  al.}, 1992) because it represents a
quantitative measure of the scaling behavior of imbalance between negative and
positive contributions in the measure.  For a positive definite measure or a
smooth field  $\kappa = 0$, while $\kappa = d/2$ for a completely stochastic 
field in a {\it d}-dimensional space (for example a field of uncorrelated 
points with $f=\pm 1$, the sign being choosen randomly and independently 
for each point, with probability~$1/2$). 
As the cancellations between negative and 
positive part of the measure decreases toward smaller scales, we get $\kappa > 
0$, and this is the interesting situation. 
It is clear that the presence of structures in the field has an 
important effect on the cancellation exponent. For example, values of 
$\kappa<d/2$, where $d$ is the dimension of the space (in the present paper
 $d=2$), indicate the presence of sign-persistent ({\it i.~e.} smooth) 
structures. 

Within turbulent flows, the value of the cancellation exponent can be related 
to the characteristic fractal dimension~$D$ of turbulent structures on all 
scales using a simple geometrical argument (Sorriso-Valvo {\em et~al.}, 2002). 
Let~$\lambda$ be the typical correlation length of that structures, of the 
order of the Taylor microscale (see for example Frisch, 1995),
so that the field is smooth (correlated) in $D$~dimensions with a cutoff 
scale~$\lambda$, and uncorrelated in the remaining $d-D$~dimensions.  
If the field is homogeneous, the partition function~(\ref{chi}) can be computed
as $(L/r)^d$ times the integral over a generic box~$Q(r)$ of size~$r$.
The scaling of the latter can be estimated integrating over regular domains 
of size~$\lambda^d$ and considering separately the number of contributions 
coming from the correlated dimensions of the field and 
those from the uncorrelated ones. The integration of the field over the 
smooth dimensions will bring a contribution proportional to their 
area~$(r/\lambda)^D$, while the uncorrelated dimensions will contribute as the 
integral of an uncorrelated field, that is proportional 
to the square root of their area~$(r/\lambda)^{(d-D)/2}$.   
Thus, when homogeneity is assumed, collecting all the contributions 
in~(\ref{chi}) leads to scaling $\chi(r)\sim r^{-(d-D)/2}$ for the 
partition function, so that one can recover the simple relation 
\begin{equation}
\kappa=(d-D)/2 \; . 
\label{kappa}
\end{equation}

\begin{figure}
\epsfxsize=12cm
\epsfbox{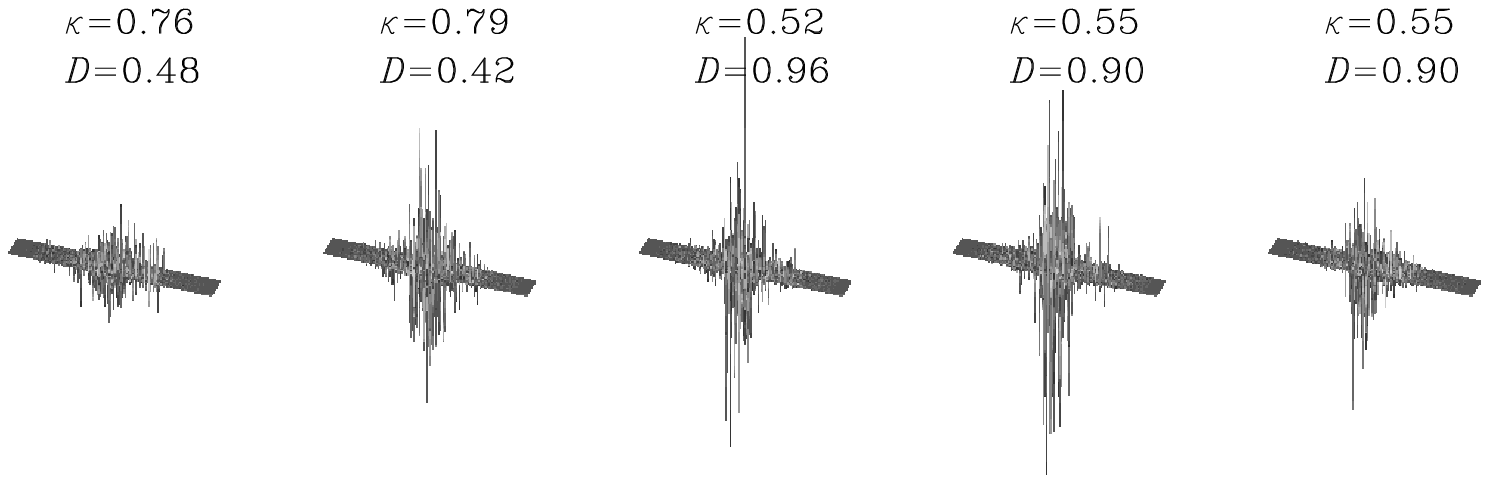}
\caption{The current helicity $H_c$ measured for the active region 
NOAA 7590. The five different plots are taken at different times, and include 
a $50$ hours interval. The presence of positive and negative strucutres
on all scales is evidenced. The flat portion of field
near the corners are due to removing in the projection effects. Over each plot
the measured cancellation exponent $\kappa$ and the fractal dimension of 
structures $D$ are reported (see further for explaination).}
\label{fig_hc}
\end{figure}

\begin{figure}
\epsfxsize=12cm
\epsfbox{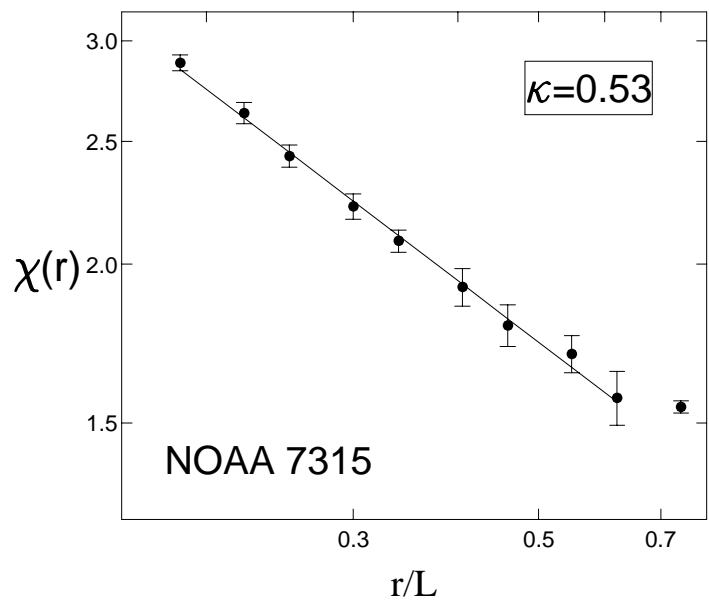}
\caption{The scaling of the partition function for a flaring active 
regions (NOAA~7315), which started to flare on October~22,~1992. The power-law 
fit is indicated as a dotted line.}
\label{fig_chi_sun}
\end{figure}

To get a quantitative measure of the change of the scaling of current helicity
inside active regions, we used observations of the vector magnetic field 
obtained with the Solar Magnetic Field Telescope of the Beijing Astronomical 
Observatory (China). Measurements were recorded in the FeI~5324.19$\,$\AA $\;$
spectral line. The field of view is about $218"\times 314"$, corresponding to 
$512\times 512$ pixels on CCD. 
The magnetic field vector at the photosphere has been obtained through the 
measurements of the four Stokes parameters, and the current density $J_z(x,y)$ 
has been calculated as a line integral of the transverse field vector over a 
closed contour of dimension $1.72" \times 1.86"$ (cf. 
Yurchischin~{\em et al.}, 2000, for details). 
The current helicity $H_c = {\bf B} \cdot {\bf J}$ (where~${\bf B}$ represents 
the magnetic field and ${\bf J} = {\bf \nabla} \times {\bf B}$ the current 
density) is a measure of small scales activity in magnetic turbulence. It 
indicates the degree of clockwise or anti-clockwise knotness of the current 
density. Let us consider a magnetogram of size~$L$ taken on the solar 
photosphere of an active region, and let ${\bf B}_{\perp}(x,y)$ the observed 
magnetic field perpendicular to the line of sight 
($(x,y)$ are the coordinate on the surface of the sun). Through 
this field we can measure the surrogate of current helicity, that is 
$h_c(x,y) = B_z(x,y) J_z(x,y)$ being $J_z(x,y) = [{\bf \nabla} \times 
{\bf B}_{\perp}]\cdot\gra{\hat{e}_z}$. \callout{Figure~\ref{fig_hc}} shows the current helicity surrogate of the active region NOAA~7590 for five different times, during the flaring event. The presence of signed structures is evident, 
as well as their evolution with time. 
A signed measure can be defined from this quantity  
   \begin{equation} 
   \mu_i(r) = \int_{Q_i(r)} h_c(x,y) dx dy  \, .
   \end{equation} 
In \callout{figure~\ref{fig_chi_sun}} we show, as example, the scaling 
behavior of $\chi(r)$~vs.~$r$ for a flaring active region (NOAA~7315) which 
started to flare on October~22,~1992. 
At larger scales we find $\chi(r)\sim const.$, and this is due to the complete
balance between positive and negative  contributions. The same behavior does
not appear at smaller scales, showing that the resolution of the images is not
high enough to resolve the smallest structures. 
In the intermediate region of scales, the cancellation exponent is found 
to be~$\kappa=0.53\pm 0.09$ (Yurchischin~{\em et~al.}, 2000).

Let us consider now what happens to the fractal dimension of current 
structures~$D$ as a function of time. To this aim, we take different 
consecutive magnetograms of the same active region, and for each magnetogram, 
we compute the value of~$\kappa$ and then of~$D$ through 
relation~(\ref{kappa}). 
Note that, since cancellations in the vertical photospheric magnetic 
field~$B_z(x,y)$ itself have been found to be very small 
(Lawrence~{\em et al.},~1993; Abramenko~{\em et al.},~1998), with a 
cancellation exponent of the order of~$10^{-2}$, cancellations of the current
helicity are entirely due to the current structures.  

\begin{figure}
\epsfxsize=12cm
\epsfbox{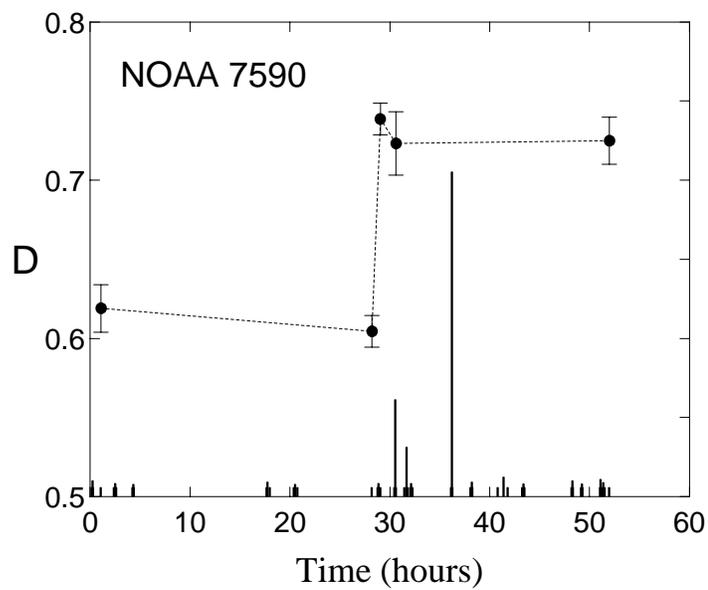}
\caption{We present one flaring event observed in~1993 (see text 
for description) in the bottom part of the plot (line bars, in arbitrary 
units). The corresponding time variation of the fractal dimension~$D$ is 
reported (symbols).}
\label{fig_Dsun}
\end{figure}

In \callout{figure~\ref{fig_Dsun}}, we report the time evolution of~$D$ 
superimposed to the flares occurred in two active regions, namely NOAAs~7315, 
and~7590 (which flared on October 1,~1993). 
Quite surprisingly we observe that the fractal dimension~$D$, 
starting from a given value ($D<1$), becomes abruptly larger in correspondence 
with a sequence of big flares occurring at the top of the active region into 
the corona. The same behavior has been
found for all calculations in all active regions we examined.  The increase
of the dimension of the structures may be the signature that dissipation
has occured. In fact, annihilation is responsible for the smoothing of the small
scales structures. 

As shown in figure~\ref{fig_chi_sun}, a saturation of~$\chi(r)$ is observed 
at large scales. At the smallest scales, the density of 
the measure must becomes smooth (no changes in sign are present) and we might 
thus found a saturation of~$\chi(r)$. The fact that we do not find this 
saturation is an indirect evidence that elementary flux tubes are smaller than
the instrumental resolution. The change towards larger~$D$ is then probabily
due to the occurrence of dissipation of smaller and smaller flux tubes, that
is magnetic energy is suddenly transferred towards small scales. This is the
occurrence of an energy cascade towards smaller scales. 

\begin{figure}
\epsfxsize=12cm
\epsfbox{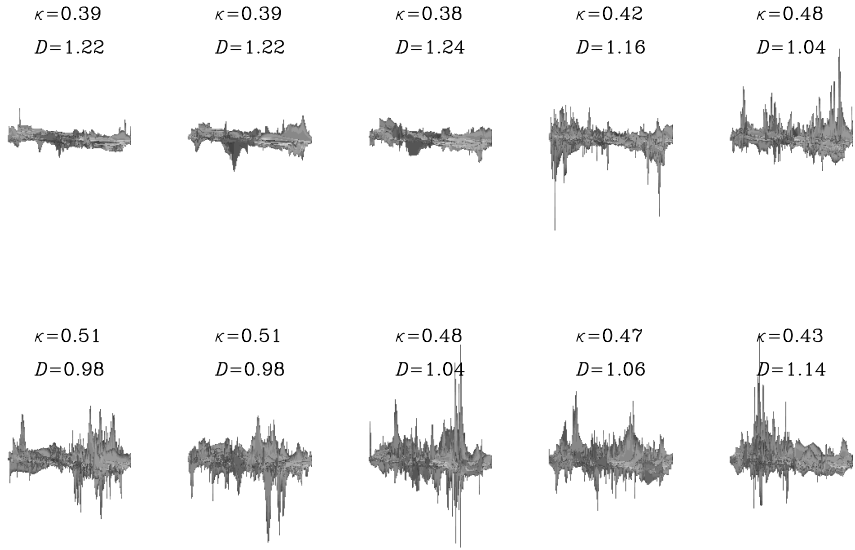}
\caption{The current field $J$ obtained from high resolution 
two-dimensional numerical simulation of MHD equations. The different 
plots are ten snapshots in the statistically steady state, from 
about $t=168$ up to $t=336$ in non-linear times units, $\tau_{NL}$. 
As for the solar data, the presence of positive and negative strucutres
on all scales is clear. We report, over each plot, the measured values of 
$\kappa$ and $D$.}
\label{fig_curr}
\end{figure}

\begin{figure}
\epsfxsize=12cm
\epsfbox{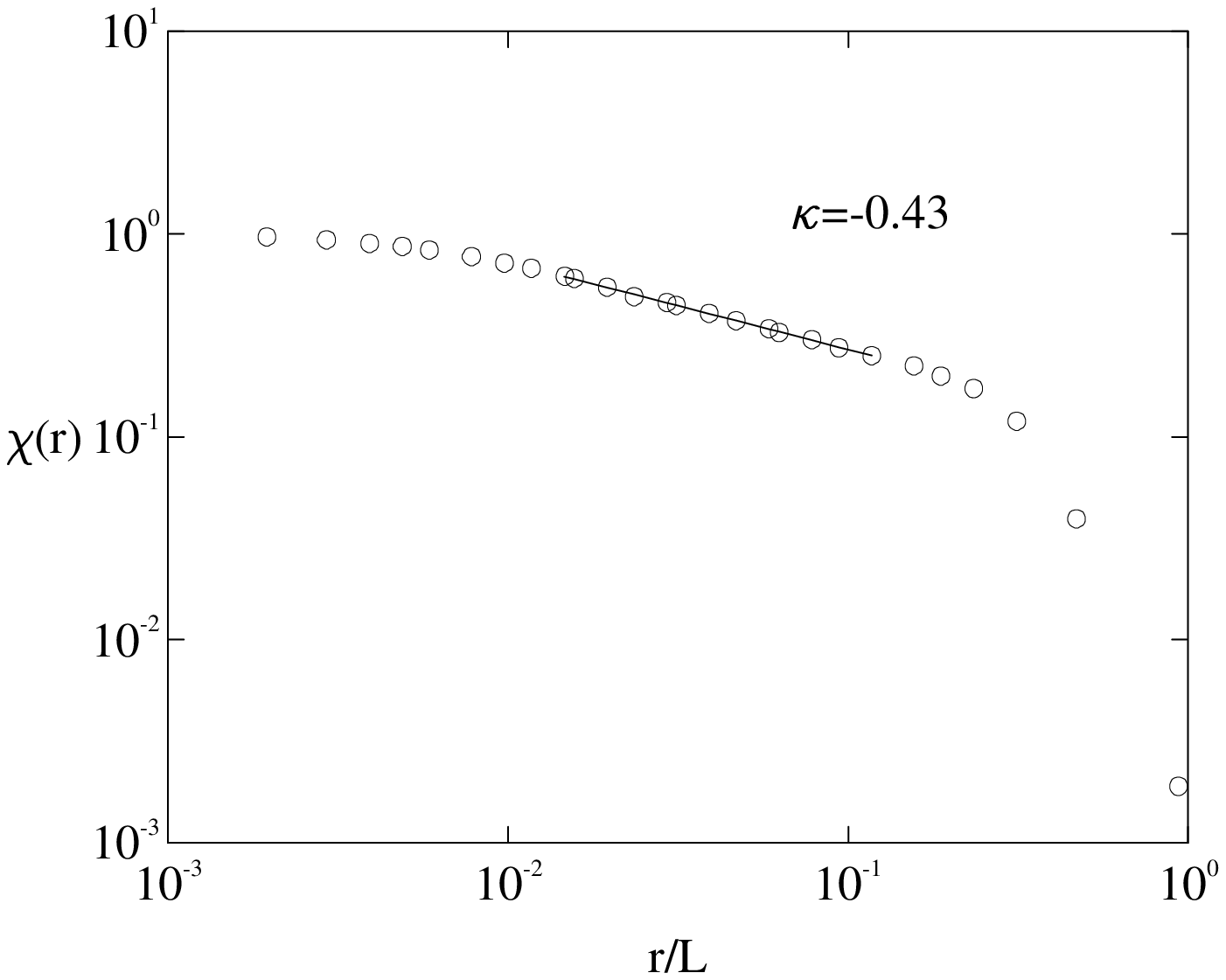}
\caption{The scaling of the partition function for the current obtained from 
numerical data. This result is obtained by averaging the time evolution, in 
order to increase the statistics. The power-law fit is indicated as a 
straight line. The scales are normalized to the simulation box size $L=2\pi$.}
\label{fig_chij2d}
\end{figure}

Using high resolution numerical simulation of two-dimensional ($d=2$) 
turbulent magnetohydrodynamic flows (Politano {\em et~al.},~1998; 
Sorriso-Valvo {\em et~al.},~2000; Sorriso-Valvo {\em et~al.},~2001), 
we can build up the signed measure for different fields. For example, since
the geometry of the magnetic field $\gra B(x,y)=(B_x,B_y,0)$ is 
two-dimensional, the current $\gra J(x,y)=\nabla\times\gra J=(0,0,J_z)$ 
has only the $z$ component, perpendicular to the 2-d simulation box,
{\it i.~e.} the plane $(x,y)$. In \callout{Figure~\ref{fig_curr}} we display 
the current field $J(x,y)$ for the numerical data, using ten snapshots in the 
statistically steady state, from $t=168$ up to $t=336$ in non-linear times 
units, $\tau_{NL}$.
As can be seen, the presence of positive and negative structures is evident as 
in the case of the solar data. A clear evolution of the complexity of the field
is present. The signed measure of the current can be then computed as usual:   
  $$
  \mu_i(r)=\int_{Q_i(r)}\, J_z(x,y)\,dx\,dy\,\, ,
  $$
and the scaling properties of the time averaged partition function are 
reported in 
\callout{figure~\ref{fig_chij2d}}. The power-law scaling~(\ref{chilaw}) 
is clearly visible in a range extending from the large scales 
(near the integral scale 
of the flow $\ell_0\sim 0.2L$, $L=2\pi$ being the size of the simulation box)
down to a correlation lenght $r^\star$ of the order of the Taylor microscale
$\lambda\sim0.02L$ of the flow (see for example Frish,~1995).  In this region,
we fit the partition function to obtain the cancellation 
exponent~$\kappa=0.43\pm0.06$. A saturation of the partition function is
observed at a scale~$r_S$ which is found to be of the order of the dissipative
scale of the flow.  In fact, for scales smaller than~$r_S$ the dissipation
stops the  structures formation cascade, so that cancellations are stopped
too.  The fractal dimension of the current structures has been computed 
using the relation~(\ref{kappa}), which gives $D\simeq 1$, 
indicating current structures similar to filaments. 
The presence of filaments can be clearly observed by a direct inspection 
of the current field contour plot, confirming the reliability of the model 
(see Sorriso-Valvo {\sl et~al.},~2002).

\begin{figure}
\epsfxsize=12cm
\epsfbox{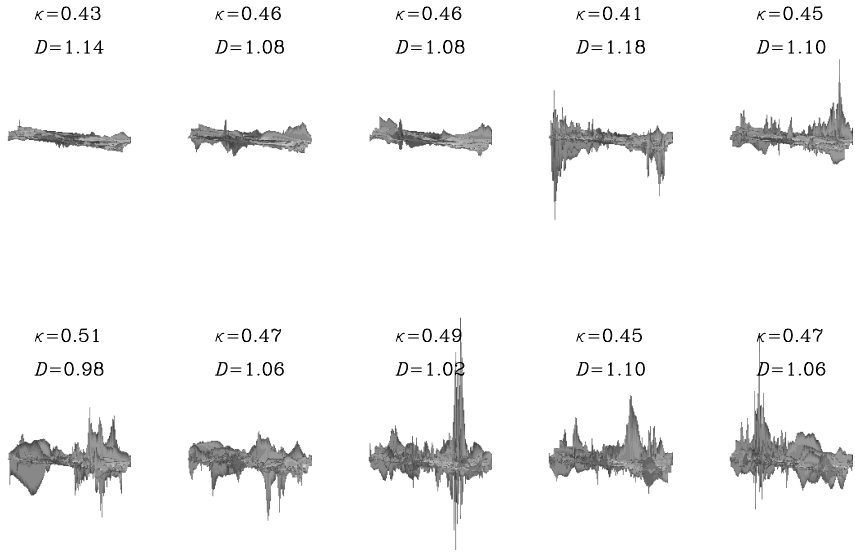}
\caption{The surrogate of the current helicity $H_c^{(2d)}$ obtained 
from high resolution two-dimensional numerical simulation of MHD equations. 
The different plots refer to the same snapshots presented for the current 
in Figure~\ref{fig_curr}. 
The signed structures are present and reproduce the current structures, 
but the field looks smoother than the current field itself.
Again, we report, over each plot, the measured values of $\kappa$ and $D$.}
\label{fig_hc2d}
\end{figure}

\begin{figure}
\epsfxsize=12cm
\epsfbox{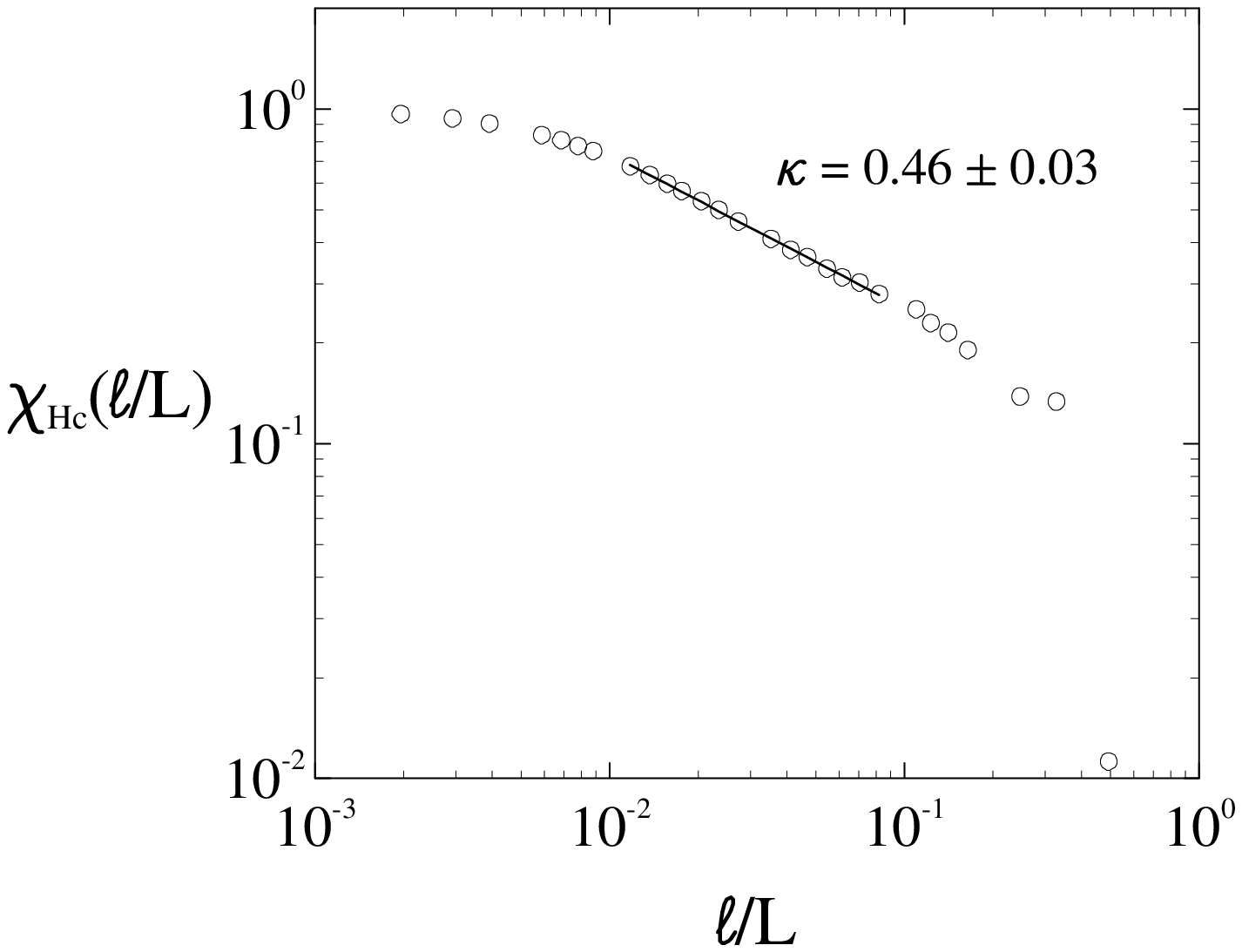}
\caption{The scaling of the partition function for the current helicity 
obtained from numerical data. As for the current, this result is obtained by 
averaging the time evolution, in order to increase the statistics. 
The power-law fit is indicated as a straight line. The scales are 
normalized to the simulation box size $L=2\pi$.}
\label{fig_chihc2d}
\end{figure}

In order to compare more directely the numerical results with the solar data 
cancellation analisys, we introduce now a new surrogate for the current 
helicity.
Since in two-dimensional geometry the current helicity is zero, the current 
being perpendicular to the magnetic field, we simply consider 
$H_c^{(2d)}(x,y)=J(x,y)|B(x,y)|$, which
is represented in \callout{Figure~\ref{fig_hc2d}} for the same times snapshots 
as in Figure~\ref{fig_curr}. The current helicity surrogate, as in the case 
of the solar data, appears smoother than the current itself, and keeps the 
same structures topology and evolution. The signed measure of such field is
then computed as in the previous cases: 
  $$
  \mu_i(r)=\int_{Q_i(r)}\, H_c^{(2d)}(x,y)\,dx\,dy\, .
  $$
The scaling properties of the partition function $\chi(r)$ can be now 
represented by the cancellation exponent, obtained by the usual fitting 
procedure after time averaging. In~\callout{Figure~\ref{fig_chihc2d}} we 
present the scaling of $\chi(r)$, together with the fitting power-law, 
for which we find an exponent $\kappa=0.46\pm 0.03$. This result is
very close to the result obtained for the current, showing that for our
numerical data as well, the current field is the one responsible for 
sign singularities, and its structures control the cancellations.

\begin{figure}
\epsfxsize=12cm
\epsfbox{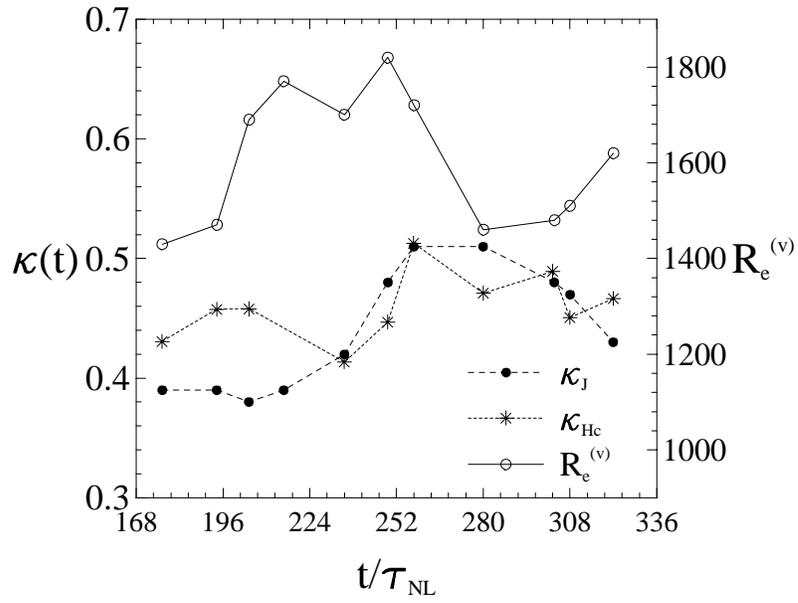}
\caption{The cancellation exponents $\kappa$ for both the current (black 
circles) and the current helicity (stars) for different times. The open circles
 plot represents the kinetic Reynolds number of the flow (left y-axis).}
\label{fig_rey}
\end{figure}

We want now to study in more detail the time evolution of the cancellations 
effects in the two-dimensional numerical simulations. To do this, we plot
in \callout{Figure~\ref{fig_rey}} the time evolution of the (kinetic) 
Reynolds number, together with the two cancellation exponents $\kappa_j$ and 
$\kappa_{H_c^{(2d)}}$, for the snapshots already presented in 
Figures~\ref{fig_curr} and~\ref{fig_hc2d}. As already pointed out, the first 
snapshots looks smoother than the others (see Figures~\ref{fig_curr} and 
\ref{fig_hc2d}, and this fact can be interpreted as stronger presence of 
dissipative effects for such times; correspondingly, the values of both the 
cancellation exponents are smaller, which means, following our model, that the
fractal dimension of the structures~$D$ is larger for these times. This
observation confirm the results already obtained in Sorriso-Valvo {\it et~al},
(2002) and discussed above.
The Reynolds number also presents a time evolution, which is however shifted 
with respect to the evolution of the cancellation exponents. Unfortunately, 
due to the limited time interval of our simulations, it is impossible to say 
whether that shift is backward or foreward. Since the Reynolds number is 
related to the importance of dissipative effects against non linear effects in 
the turbulent cascade, it would be interesting to clarify this point as a 
further confirmation of our interpretation.

In this paper we point out that the changes in the scaling behavior of
cancellations, measured through the cancellation exponent $\kappa$, are 
due to the topology changes of the structures present in the field, and are
thus related to the importance of dissipative effects. The non-linear 
turbulent cascade, underlying the formation os such structures on all scales, 
can be considered as one important input mechanism for flares. The results 
obtained from the analysis of the numerical simulations strongly support
the interpretation of the observational results for the photospheric 
magnetic field in the active regions.

To conclude, it is evident that the behavior we found can be used as a 
signature of the occurrence of big flares.  
High energy solar flares become of great interest because they can  
produce severe damages on Earth. Power blackouts, break up of communications  
and mainly damage of satellites or space flights, can be ascribed to energy  
released during big solar flares. It is then evident that the possibility of  
forecasting, even if partially, high energy flares has a wide practical  
interest to prevent the effects of flares on Earth and its environment. 
We build up a model which allows us to recognize without
ambiguity changing  behavior of the photospheric magnetic field of active
regions. These changes,  pointed out through the variation of a scaling index
for current helicity, can  be seen mainly before the eruption of big
flares. The change of  scaling index is due to the turbulent and intermittent
energy cascade towards  smaller scales, a mechanism which could be identified
as the input of flaring  activity, where energy is dissipated. The method
could allows us to forecast,  in real time, the appearence of the strongest
flaring activity above active regions.

\end{article}

\newpage

\newpage

\newpage

\newpage

\newpage

\newpage

\newpage

\newpage

\end{document}

%% file: update.tex
\makeatletter
\let\reset@font\empty
\makeatother